\documentclass[11pt]{article}
	
	\newcommand{\blind}{0}
	
    \makeatletter
    \renewcommand\section{\@startsection {section}{1}{\z@}%
                                       {-3.5ex \@plus -1ex \@minus -.2ex}%
                                       {2.3ex \@plus.2ex}%
                                       {\normalfont\fontfamily{phv}\fontsize{16}{19}\bfseries}}
    \renewcommand\subsection{\@startsection{subsection}{2}{\z@}%
                                         {-3.25ex\@plus -1ex \@minus -.2ex}%
                                         {1.5ex \@plus .2ex}%
                                         {\normalfont\fontfamily{phv}\fontsize{14}{17}\bfseries}}
    \renewcommand\subsubsection{\@startsection{subsubsection}{3}{\z@}%
                                        {-3.25ex\@plus -1ex \@minus -.2ex}%
                                         {1.5ex \@plus .2ex}%
                                         {\normalfont\normalsize\fontfamily{phv}\fontsize{14}{17}\selectfont}}
    \makeatother
	
	\usepackage{indentfirst}
	\usepackage{hyperref}
	\usepackage{changes}
	\usepackage{amsmath}
	\usepackage{graphicx}
	\usepackage{enumerate}
	\usepackage{comment}
	\usepackage{url} 
	\usepackage{amssymb} 
	\usepackage{animate} 
        \usepackage[normalem]{ulem}
        \usepackage{xcolor}
    
\begin{document}
		
		\def\spacingset#1{\renewcommand{\baselinestretch}%
			{#1}\small\normalsize} \spacingset{1}
		
		\if0\blind
		{
			\title{\bf Understanding and visualizing the statistical analysis of SN1987A neutrino data}
			\author{Marcos V. dos Santos$^a$ and Pedro Cunha de Holanda$^a$ \\
			$^a$Instituto de Física Gleb Wataghin - Unicamp }
			\date{}
			\maketitle
		} \fi
		
		\if1\blind
		{

            \title{\bf \emph{IISE Transactions} \LaTeX \ Template}
			\author{Author information is purposely removed for double-blind review}
			
\bigskip
			\bigskip
			\bigskip
			\begin{center}
				{\LARGE\bf \emph{IISE Transactions} \LaTeX \ Template}
			\end{center}
			\medskip
		} \fi
		\bigskip
		
	\begin{abstract}
          The SN1987A detection through neutrinos was an event of great importance in neutrino physics, being the first detection of neutrinos created outside our solar system, and then inaugurating the era of experimental neutrino astronomy. The data have been largely studied in many different analysis, and has presented several challenges in different aspects, since both supernova explosion dynamics and neutrino flavour conversion in such extreme environment still have many unknowns. In addition, the low statistics also invoke the need of unbinned statistical methods to compare any model proposal with data. In this paper we focus on a discussion about the  most used statistical analysis interpretation, presenting a pedagogical way to understand and visualize this comparison.
    \end{abstract}
			
	\noindent%
	{\it Keywords:} \emph{supernova}; \emph{neutrino}.

	\spacingset{1.5} 

\section{Introduction}\label{s:intro}

Particle physics provides a fertile ground to a vast number of methods to statistically compare theory and data, giving a quantitative filling in order to guide the prospects of a theory, or even reveal important accomplishments or tensions in experimental efforts.

In general, the theory can visually (and intuitively) be compared to data for most of the  methods, but one interesting phenomena, the neutrino burst detected from the supernova SN1987A, that is particularly affected by low statistics, evidence difficulties to such parallel.

In this paper we propose a visual understanding to handle this difficulty through an animation of a parameterized model and SN1987A antineutrino data.

\section{Statistical Analysis}\label{s:stat}

The quantitative comparison between theoretical predictions and experimental results is a major part of any scientific endeavour. As in many other areas, in particle physics an important quantity around such comparison is made is the detection rate of a specific event. As an example, theoretical models of solar neutrino production provide us with a steady theoretical neutrino flux. Solar neutrino experiments provides us with a detected event rate. The comparison between these two quantities can be done by translating the theoretical flux in an expected event rate, or inversely, translating the detected event rate in a compatible expected flux.

Besides, a lot of information can be extracted from the dependence of neutrino flux with its energy or time of detection. The most straightforward way to include this information on statistical analysis would be to split the total data into bins of specific energy or time intervals. Maybe the most spread statistical tool to implement these kind of analysis would be to calculate the following $\chi^2$:
\begin{equation}
    \chi^2=\sum_{i,j} (R^{th}_i - R^{ex}_i)(\sigma^{-2})_{ij} (R^{th}_j - R^{ex}_j)
\end{equation}
where the indices $i$ and $j$ track the binning of the data, $R^{th}_k$ is the theoretical prediction and $R^{ex}_k$ is the experimental data. This analysis has one great advantage: it allows to visually grasp how good is the accordance between theory and experiment in a figure where experimental data points, uncertainties and theoretical predictions can be plotted together. If theoretical predictions are contained inside the region around the experimental data points delimited by the uncertainties, than we expect a good fit. 

\begin{figure}[h]
    \centering
    \includegraphics[width=9cm]{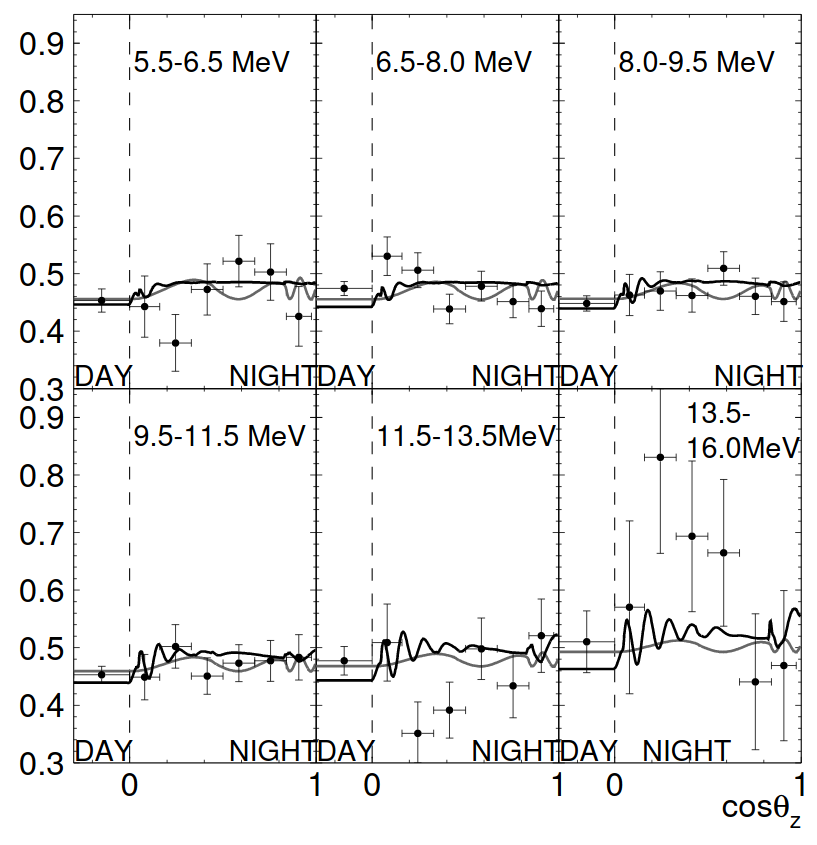}
    \caption{Data from the 1496 days of Super-Kamiokande. Each day/night plot gives a binning in energy, LOW (light gray) and LMA (black) proposed solutions to be visually compared to data points. Taken from~\cite{Smy:2002rz}.}
    \label{fig:SK1496}
\end{figure}

Several examples in neutrino physics can illustrate such procedure. Again taking solar neutrinos as an example, the data presented by the neutrino detector Super-Kamiokande is divided both in energy and time bins. 
In Figure~\ref{fig:SK1496} the data of 1496 days of running experiment are presented, and the binning on energy and time can be seen, although the binning in time is indirect, through a binning in the position of the Sun when the data was taken. The continuous line represent the prediction for the best fit point of the statistical analysis when flavour conversion is considered in two scenarios, the LMA and the LOW solutions.

As discussed, it is possible to visually get a feeling about how good the accordance between experimental data and theoretical predictions by how the theoretical curves cross the region around the experimental data within the error bars. For this particular example, two solutions to the solar neutrino problem are presented, with large mixing angle and large $\Delta m^2$ (LMA) and lower $\Delta m^2$ (LOW). We can expect by visually inspecting the figure that both solutions would fit the data quite well, fact that is confirmed by a more careful statistical analysis.

The problem of this visualization is when there is no efficient way to collect the data to form bins, for instance, due to the low event rate. In particular, when the event rate is very low it is necessary to take the experimental data event by event.

In this context, what we propose here is to recover a way to visually access how good the accordance between experimental data and theoretical prediction in a particular scenario when the statistical analysis is done event by event: the neutrino data from Supernova 1987A.

\section{Supernova 1987A}

The Core Collapse Supernova is a remarkable end of life of a star and one of the most peculiar astrophysical phenomena. Despite being a prominent optical phenomena, its most outstanding property is the powerful release of $\sim99\%$ of gravitational binding energy, generally in the order of $\sim 10^{53}$ erg, from a $m\gtrsim 10 M_\odot$ progenitor star in (anti)neutrinos of all flavors in MeV scale.

Nevertheless the high neutrino luminosity, a limitation in the neutrino observation on Earth is related to the high distance $D$ from the source, with decreasing flux proportional to $D^{-2}$, restricting the possible region for neutrino burst detection to a galactic or nearby the Milky Way, that possesses a low supernova rate of $\sim 1$ per century \cite{rozwadowska2021rate}.

\begin{figure}
    \centering
    \includegraphics[width=0.8\textwidth]{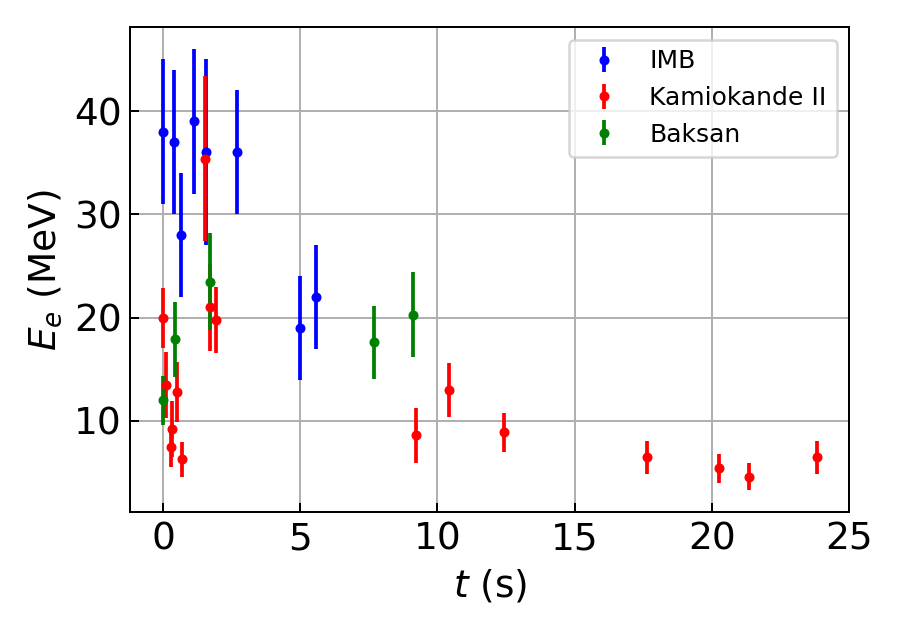}
    \caption{Positron energy and relative time from the Kamiokande II, IMB and Baksan detectors, with a total of 29 events.}
    \label{fig:data_SN1987A}
\end{figure}

Even though, in 1987, three detectors, Kamiokande II~\cite{Kamiokande-II:1987idp}, IMB~\cite{Bionta:1987qt,IMB:1987klg} and Baksan~\cite{Alekseev:1988gp},  were capable to observe a neutrino signal associated to a SN in the Large Magellanic Cloud ($\sim 50$ kpc). These data are presented in Figure~\ref{fig:data_SN1987A}. In contrast with Super-Kamiokande data in Figure~\ref{fig:SK1496}, these are individual events, and there is no obvious way to overlap a theoretical curve to them. Since the theoretical models provide a flux density and any kind of binning to convert this density into a event probability would be quite arbitrary, the approach through an unbinned maximum likelihood estimation is a robust alternative to confront the theoretical hypothesis with these individual events. In next section we describe such procedure.


\section{Modelling SN1987A event-by-event likelihood}\label{s:SN1987A-event-by-event}

Frequently the likelihood treatment in particle physics involves the usage of Poisson distribution $P(\mu, n)$, that fits well to phenomena that has small probability to occur, but a large number of tries. Given a measured variable set $\vec{x}$, the Poisson likelihood is given by:
\begin{equation}\label{eq:poissonian likelihood}
    \mathcal{L} = \prod_{i=1}^{N_{bins}} \frac{{\mu(x_i)}^{n_i}}{n_i!} e^{\mu(x_i)}
\end{equation}
where $n_i$ can be a particular number of events that occurs in a $x_i + \delta x_i$ interval of our variable, in a number $N$ of intervals, or bins, and $\mu(x_i)$ is the expected value in the same interval. It is convenient to write $\mu(x_i)$ as a distribution function on the variable $x_i$, or $\mu(x_i) = R(x_i)\delta x_i$, with a given events rate $R(x_i) = \frac{dN}{dx_i}$ in an equally spaced bin of variable $\delta x_i$ and number of counts $n_i$. Including it in equation (\ref{eq:poissonian likelihood}):
\begin{eqnarray}
    \mathcal{L} &=& \prod_{i=1}^{N_{bins}} \frac{[R(x_i) \delta x]^{n_i}}{n_i!} e^{- R(x_i) \delta x} \nonumber \\
      &=& e^{- \sum_{j=1}^{N_{bins}} R(x_j) \delta x} \prod_{i=1}^{N_{bins}} \frac{[R(x_i) \delta x]^{n_i}}{n_i!} \; .
    \label{eq:likelihood events rate binned}
\end{eqnarray}

However, binning the data to use a single expected value of a set of points requires to assume a given statistical distribution of such a bin, that generally is assumed to be Gaussian for higher number of entries. The low statistics scenario does not allow this assumption, then it is possible to model the likelihood (\ref{eq:likelihood events rate binned}) to account for each event apart. This can be made by taking the bin to an infinitesimal width $\delta x \rightarrow dx$ and number of counts $n_i \rightarrow 1$, so we consider only infinitesimal bins with one event and drop the others, then (\ref{eq:likelihood events rate binned}) becomes

\begin{equation}\label{eq:time-integrated likelihood}
    \mathcal{L} \propto  e^{- \int R(x) dx} \prod_{i=1}^{N_{obs}} R(x_i) 
\end{equation}
that also has the change from total number of bins $N_{bins}$ to total number of observed events $N_{obs}$. The idea behind maximum likelihood is to maximize the quantity in (\ref{eq:time-integrated likelihood}), or given the correspondence $\mathcal{L} = e^{-\chi^2/2}$, minimize the $\chi^2(\vec{x})=-2\log\mathcal{L}(\vec{x})$ to respect to a free set of parameters $\vec{x}$.  If we have a single event at $x=\bar{x}$, this expressions reduces to $e^{- \int R(x) dx}  R(\bar{x}) $. For different models with a normalized expected event rate $\int R(x)dx$, the likelihood is maximized for the model with the highest value of $R(x_i)$. And letting the normalization runs freely, it is maximized for $\int R(x) dx=1$. It is straightforward to note that if we consider more than one single event this maximum occurs on the total number of events.

In a supernova detection, such as SN1987A, the variables $x$ are the neutrino energy,
the detection time and events scattering angle, {\it i.e.} $R=R(t, E, \cos\theta)$. 
Then eq. (\ref{eq:time-integrated likelihood}) becomes:
\begin{equation}\label{eq:time-energy dependent likelihood}
    \mathcal{L} =  e^{- \int R(t, E, \cos\theta)\, dt\, dE \,d\cos\theta} \prod_{i=1}^{N_{obs}} R(t_i, E_i, \cos\theta_i)\, dt \,dE \,d\cos\theta
\end{equation}
where $R$ is a triply differential equation, $R= \frac{d^3N}{dt\,dE\,d\cos\theta}$ and $N$ is the expected number of events at the detector. For simplicity we did not include the scattering angle dependence on the animations presented in the following, although they were used in the  likelihood calculation. A complete analysis, including other details such as background, energy resolution and efficiency can be seen in \cite{ianni2009likelihood}, \cite{pagliaroli2009improved} and \cite{jegerlehner1996neutrino}.

\begin{figure}[htb]
    \centering
    \includegraphics[width=0.7\linewidth]{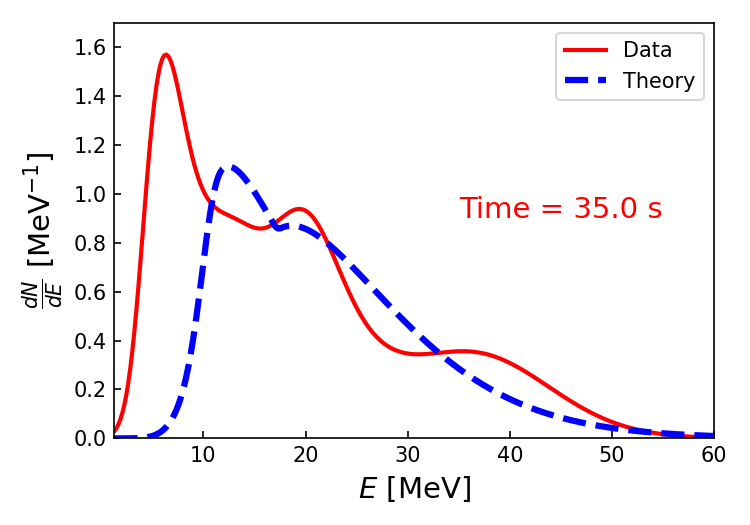}
    \caption{Theoretical events rate cumulative integrated over time (Eq. \ref{eq:events rate}) (blue) and normally distributed data as proposed in (\ref{eq:gaussian data}) (red) changing along relative detection time since the first measured neutrino from SN1987A. The time scale runs logarithmically in the first second and linearly afterwards to better show the data structure for early time events. See the animation at: {\color{blue} \url{https://github.com/santosmv/Animations-visualizing-SN1987A-data-analysis/blob/main/events_rate.png}}.}
    \label{fig:events-rate-animation}
\end{figure}


\section{Single event distribution}

The main  ingredient to construct the likelihood is the theoretical triply differential expected rate.  
However, since there is no way to convert the theoretical predictions into some quantity to be compared with individual events, we can instead modify the events to match the theoretical probability distribution. For instance, all SN1987A events are published with an uncertainty in energy, so the true information we can take from each event is a probability distribution around some most probable result. Assuming such distribution to be Gaussian, a specific event with energy $\bar{E}_\nu$ and uncertainty $\sigma_E$ on time $\bar{t}$ with uncertainty $\sigma_t$ is related to the following probability distribution:
\[
\frac{d^2P(E_\nu,t)}{dE_\nu\,dt}=\frac{1}{\sigma_E\sqrt{2\pi}}\exp{\left(-\frac{1}{2}\frac{(E_\nu-\bar{E}_\nu)^2}{\sigma_E^2}\right)}\times
\frac{1}{\sigma_t\sqrt{2\pi}}\exp{\left(-\frac{1}{2}\frac{(t-\bar{t})^2}{\sigma_t^2}\right)}
\]

This can be compared with the theoretical probability of inducing an event on the detector:
\begin{equation}\label{eq:events rate}
\frac{d^2N}{dE_\nu\,dt}=A\,\frac{d^2\phi(E_\nu,t)}{dE_\nu\,dt}\,\sigma(E_\nu)
\end{equation}
where $A$ is a proportionality factor that take into account the detector size and efficiency.

To proper visualize the data points being collected, we can create an animation with the detected event probability integrated on time. Since the uncertainty on time is very small, the distribution converges to a $\delta$-function, and such animation would advance in steps while the data gets collected:
\begin{equation}\label{eq:gaussian data}
\sum_i\frac{dP_i(E_\nu,t)}{dE_\nu}=\int_{t_0}^t dt\,\sum_i\frac{d^2P_i(E_\nu,t)}{dE_\nu\,dt}=\sum_i\frac{1}{\sigma_{E_i}\sqrt{2\pi}}\exp{\left(-\frac{1}{2}\frac{(E_\nu-\bar{E}_{\nu i})^2}{\sigma_{E_i}^2}\right)}\theta(t-t_i)
\end{equation}

Such animation is presented in Figure~\ref{fig:events-rate-animation} (red curve). Since what is presented is the cumulative result after integrating on time, the final moment of this animation, when integrated also on energy, provides all the 29 events detected by the three experiments. The comparison with theoretical predictions can be made visually if we produce a similar animation for the expected number of events, integrating Eq.~(\ref{eq:events rate}) on time, also presented in Figure \ref{fig:events-rate-animation} (dashed curve). 

The parameterization of the antineutrino\footnote{Once the detectors in 1987 were capable to measure a single channel, the inverse beta decay ($\bar{\nu}_e + p \rightarrow e^+ + n$), only electron antineutrinos could be detected.} flux $\phi(E_\nu, t)$ in Eq.~(\ref{eq:events rate}) consists in a two-component emission (accretion + cooling) and nine free parameters, that come from the proposed flux $\phi = \phi(t, E, \cos\theta, \vec{y})$, with $\vec{y} = (T_c, R_c, \tau_c, T_a, M_a, \tau_a)$, where $T_c$ ($T_a$) is the initial antineutrino (positron) temperature from the cooling (accretion) phase, $R_c$ is the radius of the neutrinosphere, $\tau_c$ ($\tau_a$) is the characteristic time from the cooling (accretion) phase and $M_a$ is the initial accreting mass \cite{pagliaroli2009improved}. The remaining three free parameters are a time offset $t_{\textrm{off}}$ to be adjusted independently for each detector.  

These parameters are estimated from an event-by-event maximum likelihood, and the best fit values of our analysis, used in Figure~\ref{fig:events-rate-animation}, are:
\begin{eqnarray}
&&T_c = 4.8 \,\mbox{MeV}, ~R_c = 6.7 \,\mbox{km}, ~\tau_c = 4.8 \,\mbox{s}, \\
&&T_a = 1.7 \,\mbox{MeV}, ~M_a = 0.4 \,M_\odot, ~\tau_a = 0.62 \,\mbox{s}
\end{eqnarray}



As described before, the maximization on the likelihood depends on two terms. The term in the exponential factor factor is related to the number of events, and drives the theoretical parameters to those who provides the right expected number of events, {\it i.e.}, the area under the curves in the end of the animation in Figure~\ref{fig:events-rate-animation}. It is quite easy to grasp if our theoretical model fits well the data by this aspect.

The second term access how close the theoretical curve is to the experimental one at the data central points, both in energy and in time. Since the uncertainty in time is negligible, we can visually compare the curves at the moments a new data is collected, providing us with a visual tool to this second ingredient of the statistical analysis. By performing these two analysis on  Figure~\ref{fig:events-rate-animation}, we can expect that, although not perfect, the theoretical prediction would provide a reasonably good fit to the data.
\begin{figure}[ht]
    \centering
    \includegraphics[width=0.7\linewidth]{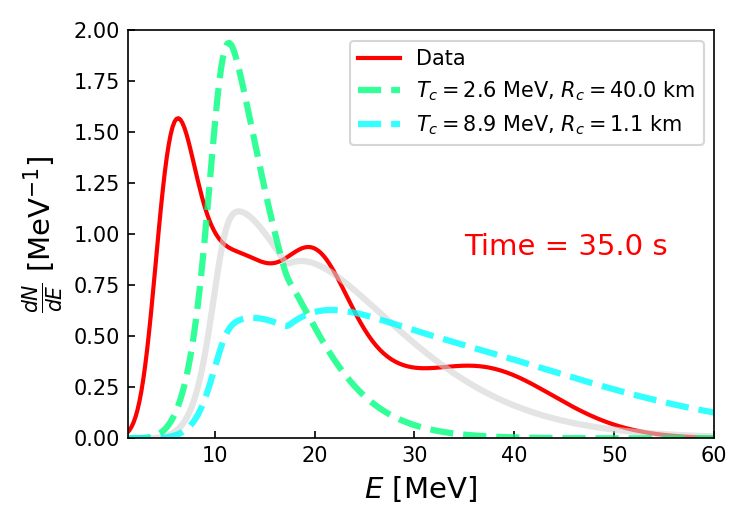}
    \caption{To visualize the effect of spectral distortion impact on events rate, we used two sets of parameters for $T_a$ and $M_a$  (in green and in cyan) that are excluded at $90\%$ C.L. according to our analysis. The green curve produces a distortion for low energy events, while the cyan produces a distortion that favour high-energy events. Also on light grey we present the best-fit point of our analyses, shown in Fig.~\ref{fig:data_SN1987A}. See the animation at: {\color{blue} \url{https://github.com/santosmv/Animations-visualizing-SN1987A-data-analysis/blob/main/events_rate_worse_fit.png}}}
    \label{fig:events-rate-animation2}
\end{figure}

It is useful now to analyse a set of theoretical parameters that do not fit well the data. This is done in Fig~\ref{fig:events-rate-animation2}, where we chosed two set of parameters that are excluded at 90\% C.L. according to our analysis. 
These parameters were chosen in a way to not change the total number of predicted events, so we can focus on the energy spectrum information. It is clear, again using  a visual comparison, that these new set of parameters produce a worse fit to the data, fact that is confirmed by a full statistical analysis.

As it was pointed out earlier, the two main neutrino observables that we are taking into consideration are the neutrino energy and the time of detection. After discussing the first on the above analysis, we will now focus on the second, and the best way to do this is the  limits in neutrino mass that can be achieved using this statistical method.

\section{Neutrino mass limits}

An important remark is that the neutrino detection spread in time is an important source of physical information, allowing us to probe both Supernova explosion mechanisms and neutrino properties. The most important neutrino property that can be probed by such time spread is its mass.


The first difficulty in these kind of analysis is that the data itself does not allow us to correlate the time of arrival of the neutrino burst at the detectors with the unknown time at which the neutrinos left the Supernova. The solution is to use the data itself to establish, through statistical analysis, the match between the neutrino flux theoretical prediction and the data, taking the time of arrival of the first neutrino event in each detector as a marker. The time of the following events, $t_i$ are taken as relative ones to the time of arrival of the first event, $t_1$:
\[
\delta_i = t_i - t_1
\]
and $t_1$ is left to vary freely to best match the theoretical prediction in a previously established time scale.

This simple picture arises when we assume massless neutrinos. In this case the relative time between events is identical to the relative time between the emission of these detected neutrinos on the production site, since the time delay due to the travel between the supernova and the detectors does not depend on the neutrino properties. In Fig.~\ref{fig:events-rate-animation} it is assumed a vanishing mass neutrino, and the time showed on the animation correspond to the time since the supernova offset.

\begin{figure}[htb]
    \centering
    \includegraphics[width=0.7\linewidth]{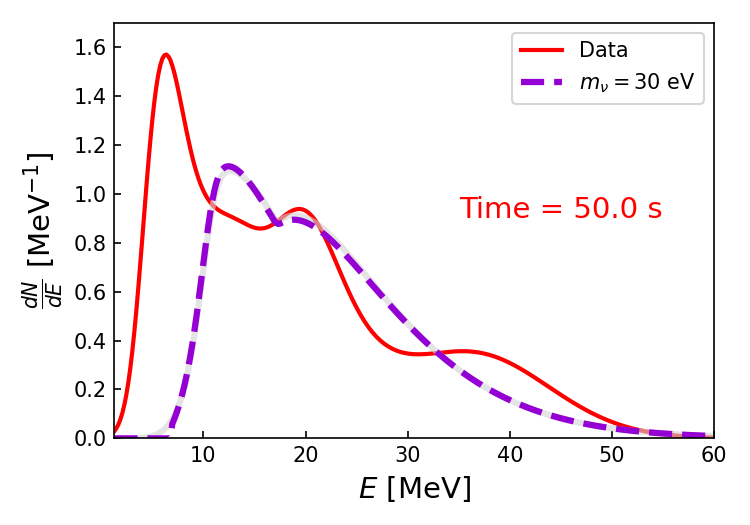}
    \caption{Effect of neutrino mass delay on SN1987A detected burst compared to standard flux for a $3 \sigma$ excluded neutrino mass. The gray line corresponds to the fitted theory in Fig. \ref{fig:events-rate-animation}. See the animation at: {\color{blue} \url{https://github.com/santosmv/Animations-visualizing-SN1987A-data-analysis/blob/main/events_rate_mass_delay.png}}.}
    \label{fig:events-rate-animation-mass-delay}
\end{figure}

However, since the neutrinos have mass, neutrinos with different energies have different velocities, which changes the described scenario. More energetic neutrinos travels faster then less energetic neutrinos, meaning that the relative times between events does not correspond to relative times of the neutrinos emission. The correction is done by a simple kinematic analysis:
\[
t_{i.d}=t_{i,p}+\frac{D}{v_i}=t_{i,p}+\frac{D}{c}\sqrt{1-\frac{m^2}{p_i^2}}
\sim t_{i,p}+\frac{D}{c}\left(1-\frac{m^2}{2E_i^2}\right)
\]
where $D$ is the distance to the supernova, and $m$ and $E$ are the neutrino mass and event energy. The sub-index $p$ ($d$) refers to the time at production (detection). The emission time of each event is then calculated from the relative times $\delta_i$, and the kinematic corrections:
\begin{eqnarray}
t_{i,d}&=&t_{1,d}+\delta_i \nonumber \\
t_{i,p}&=&\delta_i+\left(t_{i,p}-\frac{D}{c}\frac{m^2}{2E_1^2}\right)+\frac{D}{c}\frac{m^2}{2E_i^2}
\end{eqnarray}
For more details, we refere to~\cite{pagliaroli2009improved,Pagliaroli:2010ik}. 

Instead of making the correction on the time of the production, presented here to give proper credit to the authors that proposed and performed this analysis, we prefer to correct the theoretical predictions by continuous spread in time on the neutrino flux spectrum at the detector. So, instead of converting the time of the detected events to the supernova emission, we adjust the theoretical prediction to the detector site. Clearly both choices are equivalent, but with this second procedure we can use the same data animation presented in Fig.~\ref{fig:events-rate-animation}, and adjust the theoretical curve by making the replacement:
\[
t\rightarrow t-\frac{D}{c}\frac{m^2}{2E^2}
\]
in Eq.~(\ref{eq:events rate}).



 An animation evidencing this model dependent limit is shown in Fig.~\ref{fig:events-rate-animation-mass-delay}, for a  $30$ eV neutrino mass, and the same astrophysical parameters used to produce Fig.~\ref{fig:events-rate-animation}, and then with the same neutrino flux at the source. But due to the different time lag of neutrinos traveling to Earth with different energies, the time history of the expected number of events changes significantly, allowing us to place a limit on neutrino analysis using a proper statistical analysis.

\section{Conclusion}\label{s:conclusion}

This paper intended to present a pedagogical view of how to understand the likelihood analysis when an event-by-event treatment is necessary. The detection of SN1987A is a perfect example for that, once a lot of physics can be extracted by the few events that were collected through neutrino detection. It also has the interesting feature that different information can be extracted from the total expected number of events, its spectral distortion or its time structure. We present some animations as a visual tool to understand the statistical procedure, and produce a first impression on how different models fit the data.

\if0\blind{
\section*{Acknowledgements}

 This study was financed in part by the Coordenação de Aperfeiçoamento de Pessoal de Nível Superior - Brasil (CAPES) - Finance Code 001. The authors are also thankful for the support of FAPESP funding Grant 2014/19164-6. The authors are thankful to Pedro Dedin for usefull discussions during the production of this article.} \fi

\bibliographystyle{unsrt}
\spacingset{1}
\bibliography{references}
	
\end{document}